\documentclass[%
 aip,
 amsmath,amssymb,
 reprint,%
]{revtex4-1}

\usepackage{graphicx}
\usepackage{dcolumn}
\usepackage{bm}

\usepackage[utf8]{inputenc}
\usepackage[T1]{fontenc}
\usepackage{mathptmx}
\usepackage{textcomp}
\usepackage{newunicodechar}
\newunicodechar{̈}{\"{}}
\usepackage{etoolbox}
\usepackage[version=3]{mhchem} 
\usepackage{mathrsfs} 
\usepackage{amsmath,amssymb}
\usepackage{siunitx}
\usepackage{tcolorbox} 

\makeatletter
\def\@email#1#2{%
 \endgroup
 \patchcmd{\titleblock@produce}
  {\frontmatter@RRAPformat}
  {\frontmatter@RRAPformat{\produce@RRAP{*#1\href{mailto:#2}{#2}}}\frontmatter@RRAPformat}
  {}{}
}%
\makeatother
\begin{document}

\preprint{AIP/123-QED}

\newcommand*\mycommand[1]{\texttt{\emph{#1}}}

\title{Position-Dependent Calibration and Frequency Stability in On-Axis Optical Transduction of Vertical InP Nanowire Resonators}

\author{Robert G. West}
\affiliation{Institute of Sensor and Actuator Systems, TU Wien, 1040 Vienna, Austria} 
\author{Kostas Kanellopulos}
\affiliation{Institute of Sensor and Actuator Systems, TU Wien, 1040 Vienna, Austria} 
\author{Lukas Hrachowina}
\affiliation{NanoLund and Division of Solid State Physics, Lund University, 221 00 Lund, Sweden} 
\author{Magnus Borgstr\"om}
\affiliation{NanoLund and Division of Solid State Physics, Lund University, 221 00 Lund, Sweden}
\affiliation{Wallenberg Initiative Materials Science for Sustainability, Department of Solid state physics, Lund University, 221 00 Lund, Sweden.}
\author{Silvan Schmid}
\email{silvan.schmid@tuwien.ac.at}
\affiliation{Institute of Sensor and Actuator Systems, TU Wien, 1040 Vienna, Austria}

\date{\today}

\begin{abstract}
We present a quantitative framework for on-axis optical transduction of vertical InP nanowire resonators, correlating laser position to signal amplitude, calibration, and frequency stability. Photothermal resonance detuning is used to reconstruct the local beam intensity profile and to calibrate the photodetector signal using the thermomechanical noise. A noise model incorporating shot noise and spatial variation in substrate reflectance predicts the position-dependent Allan deviation. We find that the optimal detection position lies near the steepest intensity gradient, and that increasing laser power does not significantly improve frequency stability, because the accompanying temperature rise enhances thermomechanical noise and offsets the signal gain. These results establish design guidelines for optimizing nanowire-based sensors in on-axis optical detection schemes.
\end{abstract}

\maketitle

Owing to their extremely low mass and high aspect ratio, nanomechanical resonators exhibit strong frequency shifts in response to minute changes in mass,\cite{ekinci2004ultrasensitive,hanay2012single} temperature,\cite{schmid2023fundamentals} and force,\cite{bachtold2022mesoscopic} making them attractive for mass spectrometry,\cite{sader2024data,neumann2024hybrid} force sensing,\cite{eichler2022ultra} bolometry and calorimetry,\cite{kim2021nanomechanical} spin-state detection,\cite{teissier2014strain,ovartchaiyapong2014dynamic} and photothermal spectroscopy.\cite{west2023photothermal}

Among nanomechanical resonators, nanowires (NWs) stand out for their exceptionally low mass and large aspect ratio, which enable ultrasensitive detection in the attogram-to-zeptogram regime for mass\cite{gil2010nanomechanical,gruber2019mass,escobar2023nanomechanical} and the zeptonewton regime for force.\cite{Hallstrom2010fifteen,moser2013ultrasensitive,de2017universal,braakman2019force,fogliano2021ultrasensitive,bachtold2022mesoscopic} NWs also offer a high dynamic range exceeding 90~dB in the linear regime.\cite{postma2005dynamic,nichol2009controlling,molina2021dynamic} 

The mechanical motion of NWs can be transduced electronically,\cite{li2007ultra} acoustically (for pillars),\cite{kaehler2023transduction} by scanning tunneling microscopy,\cite{fian2010new} and optically. Optical methods include interferometry,\cite{belov2008mechanical,nichol2008displacement}
Bragg scattering,\cite{houlton2018optical}
stroboscopic imaging,\cite{hessman2007high}
and modulated light scattering, which enable transduction even of densely packed NW arrays.\cite{poot2012mechanical,fountaine2016near,houlton2018optical,hrachowina2021wafer,doster2022observing} While some optical configurations illuminate the NW perpendicular to its long axis, on-axis (parallel) illumination is more straightforward, provides symmetric transduction of degenerate modes, and offers greater flexibility for simultaneous measurements of multiple NWs.\cite{gil2010nanomechanical,yang2013multimode,molina2020optical,molina2021dynamic,kaehler2023transduction} 

The on-axis readout scheme benefits from strong scattering by the metal catalyst particle at the NW tip — an inherent byproduct of the growth process.\cite{molina2020optical} In the case of indium phosphide (InP) NWs, the large absorption cross-section provides an additional enhancement of the optomechanical interaction, though this is not generally expected for other NW materials.\cite{anttu2014absorption} The reflected-light signal at the photodetector is strongly position-dependent, shaped by the laser intensity profile and the position-dependent substrate reflectance. Although NW resonators have been extensively studied, a quantitative framework linking laser position to transduction efficiency, noise performance, and frequency stability in on-axis optical detection remains lacking. Here, we use vertical InP nanowires, previously unstudied as mechanical resonators, to establish such a framework.

However, the same NW properties that enable high sensitivity also introduce complexity. Optomechanical and photothermal back-actions, as well as differential expansion gradients, can induce signal instability and frequency drift, requiring careful control of laser power and detection position.\cite{bellon2024temperature,amusia2020viscous,hossain2019} Moreover, NW material properties (i.e. Young's modulus, thermal conductivity, and internal friction) vary strongly with diameter, surface state, and crystal quality,\cite{evoy2000temperature,dos_Santos_2010,Wang2017Nanowires,carrete2009molecular,matthews2012heat} necessitating per-device characterization of the transduction chain.

While InP NWs are well characterized optically and electronically, and have hosted embedded quantum dots for spin-based studies,\cite{vanWeert2009selective,montinaro2014quantum} their mechanical resonances have not been widely exploited. InP is attractive for nanomechanical sensing due to its established epitaxial growth and compatibility with III--V heterostructure integration; however, its strong optical absorption can introduce excess photothermal effects and associated noise.\cite{anttu2014absorption} 
The InP NWs used in this study were grown on an InP substrate by metal--organic vapor phase epitaxy (MOVPE) in the particle-assisted growth mode using Au nanoparticle catalysts, resulting in growth along the $[\bar{1}\bar{1}\bar{1}]$ direction. In contrast to the lithographically patterned arrays reported in Ref.~\citenum{hrachowina2021wafer}, the Au seeds here were deposited from an aerosol, producing randomly positioned NWs with an average surface density of $0.04~\mu\mathrm{m}^{-2}$. The resulting NWs have lengths of $L\approx14\,\mu$m, tip diameters of ${\sim}107$~nm, and slightly tapered bases of ${\sim}150$~nm.

\begin{figure}
\centering
\vspace{-3mm}
\includegraphics[width=0.5\textwidth]{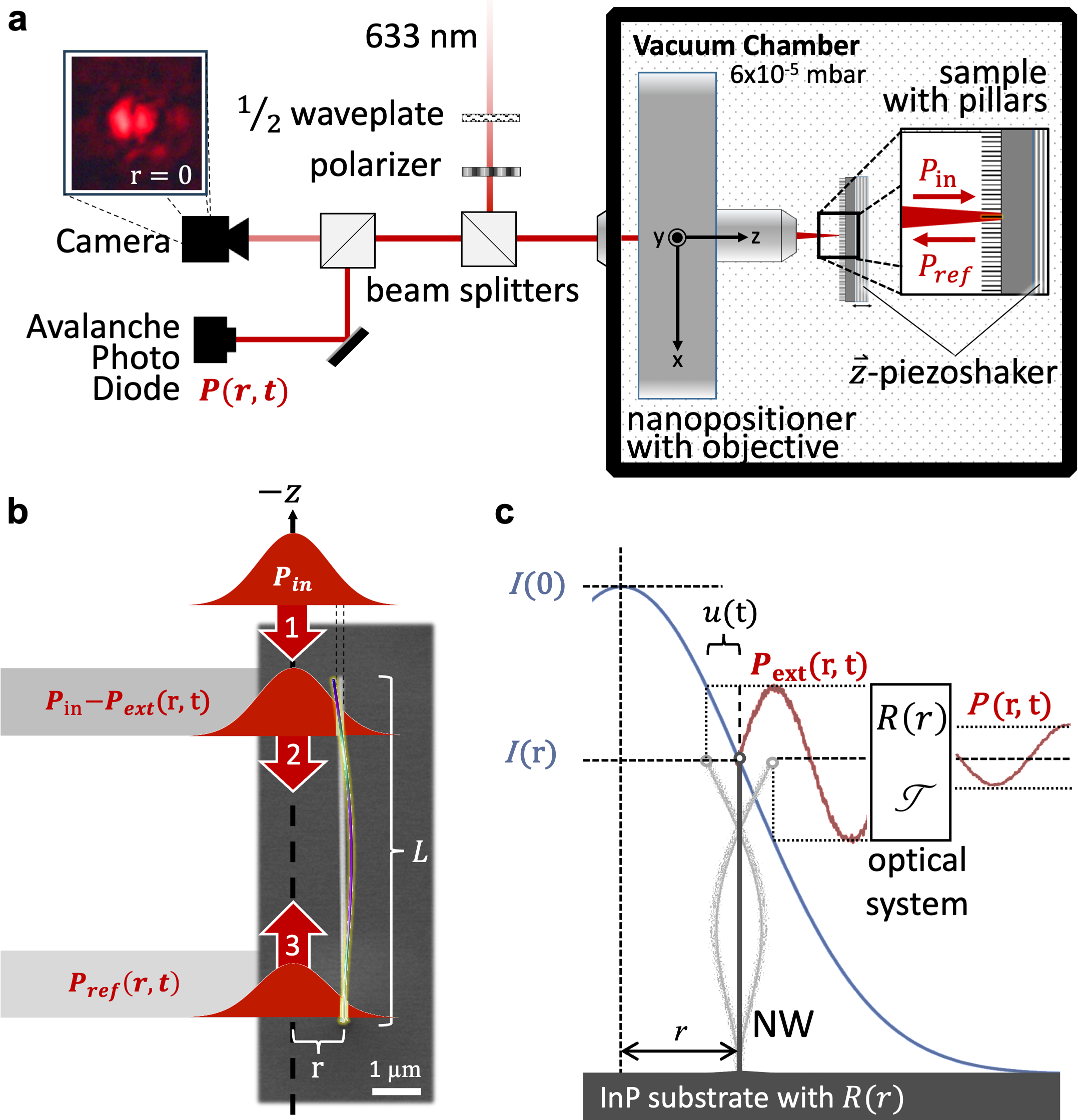}
\caption{(a) Schematic of the optical setup: a laser is focused on an \ce{InP} nanowire, and the reflected light, modulated by the nanowire's motion, is detected by the photodetector (PD). The PD signal is processed by a lock-in amplifier with a PLL, while a piezoshaker drives the nanowire from behind. (b) SEM image of a nanowire with an overlaid exaggerated FEM-simulated second harmonic mode shape; 
The incident laser power ($P_\mathrm{in}$) is scattered and partially absorbed by the nanowire ($P_\mathrm{ext}$), with the remainder reflected to the PD ($P_\mathrm{ref}$). (c) Illustration of how the nanowire's motion modulates the local laser intensity profile.}
\label{fig:schematic}
\end{figure}

The on-axis optical transduction concept, depicted in Fig.~\ref{fig:schematic}a, is similar to that employed by Molina et al.,\cite{molina2020optical,molina2021dynamic} with the exception of a higher-magnification (50x) objective, which yields a smaller beam waist and sharper intensity gradient. The NW resonant motion is transduced using a 633~nm laser aligned with the NW's long axis, with a beam waist $w_0$ of 992~nm as determined by the knife-edge method (see supplementary material).

The NWs are actuated from below the substrate by a piezoelectric actuator in vacuum ($<6\times10^{-5}$~mbar; Fig.~\ref{fig:schematic}a). The laser position along the axis ($r$) of flexural motion is controlled by mounting the objective on a 3-axis nanopositioner. The total reflected light $P_\mathrm{ref}(r,t)$ (Fig.~\ref{fig:schematic}b) equals the incident power $P_\mathrm{in}$ minus the power extinguished by absorption and scattering $P_\mathrm{ext}(r,t)$ at the NW position, assuming complete absorption of all light within the NW's absorption cross-section before it reaches the substrate.\cite{anttu2014absorption} Any systematic reduction of the remaining light before re-entering the objective, including reflection by the bulged InP substrate, is captured by a reflectance $\mathrm{R}(r)$. The light then passes through the optical system with transmittance $\mathcal{T}$ before reaching the PD with a power
\begin{equation}\label{eq:P}
\begin{split}
P(r,t) &= \mathcal{T}R(r)[ P_\mathrm{in} - P_\mathrm{ext}(r,t) ].
\end{split}
\end{equation}

NW motion modulates the local optical intensity within the Gaussian beam profile (Fig.~\ref{fig:schematic}c), and the signal is maximized by positioning the NW at the steepest intensity gradient $dI(r)/dr$, located at half the beam waist $r=w_0/2$. Therefore, the voltage output of the AC-coupled photodetector is scaled by this gradient times the oscillation amplitude $u(t)$, the effective extinction coefficient of the NW $\sigma_{\mathrm{ext}}$, and the photodetector gain $g_{\mathrm{PD}}$. Since the oscillation amplitude $u(t) \ll w_0$, the Gaussian slope can be linearized, yielding the voltage modulation
\begin{equation}\label{eq:VAC}
\begin{split}
V_\mathrm{AC}(r,t)
&= g_{\mathrm{PD}}\mathcal{T}R(r) \;\frac{\mathrm{d}I(r)}{\mathrm{d}r} \sigma_\mathrm{ext} \; u(t)\\
& = [1/C(r)] \; u(t)
\end{split}
\end{equation}
with the extinction cross-section $\sigma_{\mathrm{ext}}$ of the NW, in units [m$^2$], the irradiance $I(r)$, in units [Wm$^{-2}$], and the calibration factor $C(r)$, in units [mV$^{-1}$]. 
The latter absorbs all unknown parameters of the optical readout chain and can be evaluated by measuring a known NW displacement, such as its thermomechanical noise.\cite{schmid2023fundamentals}
According to (\ref{eq:VAC}), it can be obtained by comparing the measured voltage power spectral density (PSD), $S_{V_\mathrm{thm}}$, to the theoretical thermomechanical displacement noise PSD, $S_{u_\mathrm{thm}}$, at resonance
\begin{equation}
C(r) = \sqrt{\frac{S_{u_\mathrm{thm}}(\omega_0)}{S_{V_\mathrm{thm}}(\omega_0, r)}}\,,
\label{eq:Cc}
\end{equation}
with $\omega_0$ being the resonance angular frequency. The on-resonance thermomechanical displacement noise PSD is given by \cite{schmid2023fundamentals}
\begin{equation}\label{eq:Suth}
S_{u_\mathrm{thm}}(\omega_0)=\frac{4k_\mathrm{B}TQ}{m_\mathrm{eff}\omega_0^3},
\end{equation}
where $T$ is the mean NW temperature, $Q$ is the quality factor, and $m_{\mathrm{eff}}$ is the effective mass. While $Q$ can be measured and $m_{\mathrm{eff}}$ can be calculated, the NW temperature must be estimated.

\begin{figure}
\centering
\includegraphics[width=0.465\textwidth]{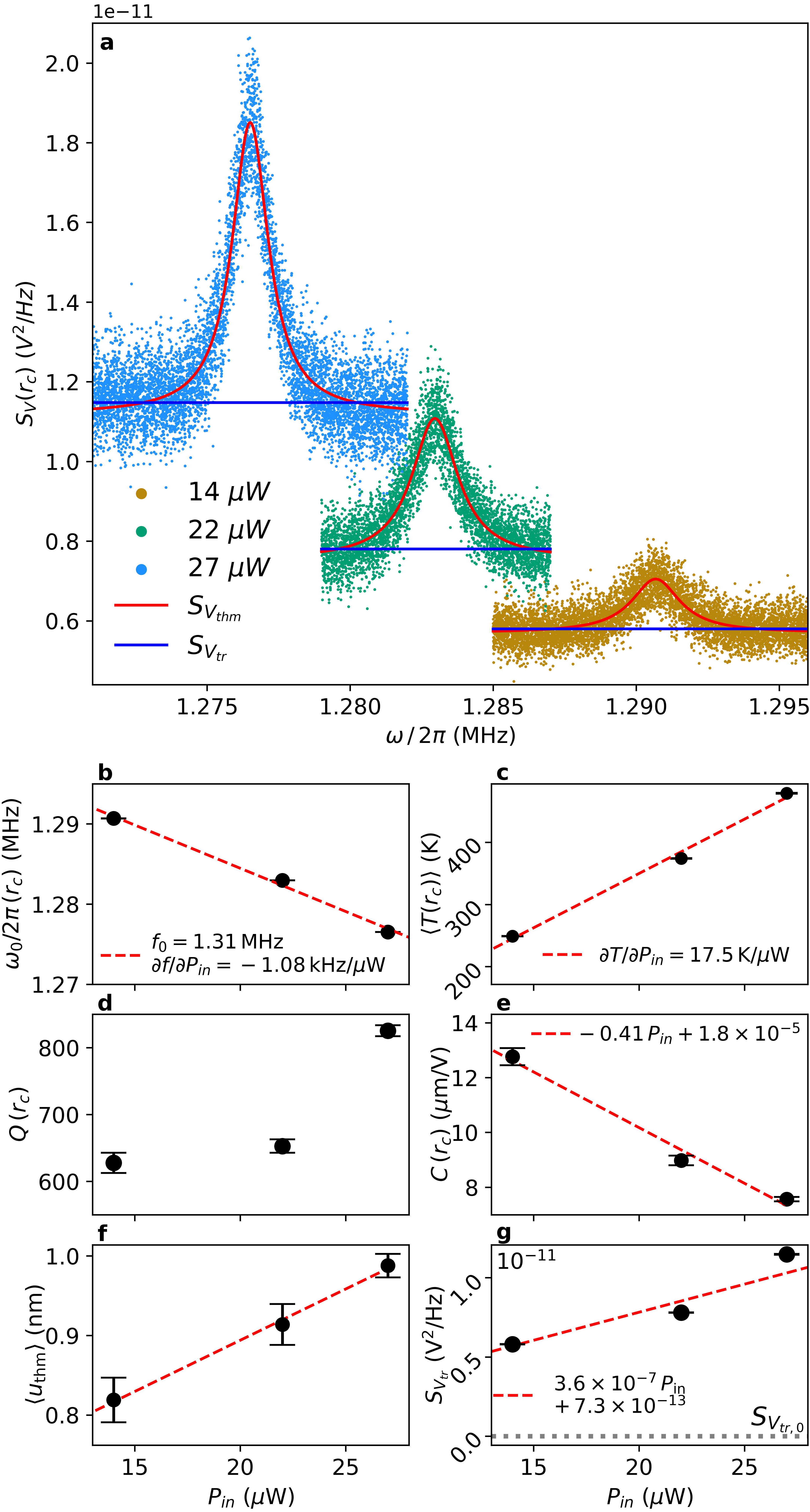}
\caption{(a) Power spectral densities (PSDs) of the thermomechanical signal measured at
the point of maximum sensitivity ($r_c$) for three incident laser powers, each fitted with a Lorentzian (red). The blue line indicates the background transduction noise floor. 
(b)~Resonance frequency shift with a linear fit. 
(c)~Mean temperature rise estimated from FEM simulations.
(d)~Quality factor. 
(e)~Calibration factor with a linear fit (see Eq.~\ref{eq:Cc}). 
(f)~Inferred mean thermomechanical amplitude $u_\mathrm{thm}$. 
(g)~Background transduction noise $S_{V_{\mathrm{tr}}}$ with a
linear fit; the gray dashed line indicates the $P_\mathrm{in} = 0$ baseline.
All data in (b)--(g) are plotted as a function of the incident optical power $P_\mathrm{in}$.}
\label{fig:calibfig}
\end{figure}

Fig.~\ref{fig:calibfig}a shows the measured voltage PSD of the NW's thermomechanical noise at resonance for different incident powers $P_\mathrm{in}$. Each PSD is decomposed into a Lorentzian resonance component ($S_{V_\mathrm{thm}}(\omega,r)$, red lines) and a residual transduction noise floor ($S_{V_\mathrm{tr}}(\omega,r)$, blue lines). 

The NW resonance frequency shifts downward with laser power, suggesting a temperature-induced material softening (Fig.~\ref{fig:calibfig}b). The rate of approximately $1.1~\mathrm{kHz}/\mu\mathrm{W}$ yields an extrapolated zero-power resonance frequency of $\omega_0 \approx 2\pi \cdot 1.31~\mathrm{MHz}$. The associated mean temperature rise $\langle\Delta T\rangle$ was inferred from the frequency shift using a finite-element simulation with bulk material parameters (Fig.~\ref{fig:calibfig}c). Fixing the substrate temperature and varying that of the NW yielded a temperature responsivity of $\mathscr{R}_T\approx-47.1~\mathrm{ppm\,K^{-1}}$, corresponding to temperature rise of $\approx17.5~\mathrm{K}/\mu\mathrm{W}$. 

This temperature increase is primarily driven by the strong absorption of the InP NWs (${\sim}94\%$), consistent with the observations reported by Anttu et al.~\cite{anttu2014absorption} under similar conditions. Assuming a thermal conductivity of $9~\mathrm{W\,m^{-1}K^{-1}}$,\cite{mingo2004lattice} we estimate the NW absorption cross-section as $\sigma_\mathrm{abs}\approx2.0\times10^{-13}~\mathrm{m}^2$, approximately $20\times$ larger than its geometric cross-section. For comparison, the plasmonic absorption cross-section of the hemispherical Au tip, estimated via Mie theory, is $\approx9\times10^{-16}~\mathrm{m}^2$ — about $200\times$ smaller than that of the NW.

Substrate-mediated heating is neglected, in agreement with the results of Anttu et al., who reported only $\mu\mathrm{K}$-scale contributions even for shorter NWs ($L\approx4.5\,\mu\mathrm{m}$).

We further observe the onset of anomalous behavior at $27$~µW, while higher powers ($\gtrsim40~\mu$W) lead to irreversible damage. These thresholds are consistent with reported values of ${\sim}70$--100~$\mu$W for ${\sim}100$~nm InAs NWs.\cite{he2012laser}

For the NW dimensions considered here, the thermal relaxation time is estimated to be $\approx 35\,\mu\mathrm{s}$ for a molar heat capacity of $46~\mathrm{J\,mol^{-1}K^{-1}}$,\cite{vasil2006thermodynamic} significantly longer than the oscillation cycle.

The quality factor $Q$ increases with power (Fig.~\ref{fig:calibfig}d). Since the thermal
time constant is a lot longer than the oscillation period, possible photothermal back-action amplification of the NW vibration can be neglected. The most likely explanation for the observed increase of $Q$ is the thermally induced reduction of surface losses. Due to the large surface-to-volume ratio of the nanowire, surface dissipation is expected to dominate the total mechanical loss.\cite{villanueva2014evidence} At elevated temperatures, surface adsorbates and native oxide defect states on InP likely desorb or restructure, reducing the density of dissipative surface states and thus the overall damping.\cite{tao2015permanent}

With known $T$ and $Q$, and an effective mass of $m_{\mathrm{eff}}=$ 170~fg (see supplementary material), (\ref{eq:Suth}) yields the thermomechanical noise, from which the calibration factor (\ref{eq:Cc}) follows directly (Fig.~\ref{fig:calibfig}e). $C$ decreases with increasing laser power, as expected from (\ref{eq:VAC}). The inferred mean thermomechanical NW amplitude \cite{schmid2023fundamentals} $\langle u_\mathrm{thm}\rangle=\sqrt{k_B T/(m_{\mathrm{eff}}\omega_0^2)}$ increases linearly with power (Fig.~\ref{fig:calibfig}f), driven primarily by the corresponding linear increase in temperature.

Finally, the transduction noise floor rises nearly linearly with power (Fig.~\ref{fig:calibfig}g), suggesting that the measurement transduction is shot-noise-limited. Extrapolating to zero power gives an intrinsic electronic noise level of $S_{\mathrm{V_{tr,0}}}\approx7.3 \times10^{-13}~\mathrm{V^2/Hz}$ (dashed line), on the same order of magnitude as the typical intrinsic noise floor of the detector, indicating that the measurement is predominantly detector-limited.

\begin{figure*}
    \centering
    \includegraphics[width=0.9\textwidth]{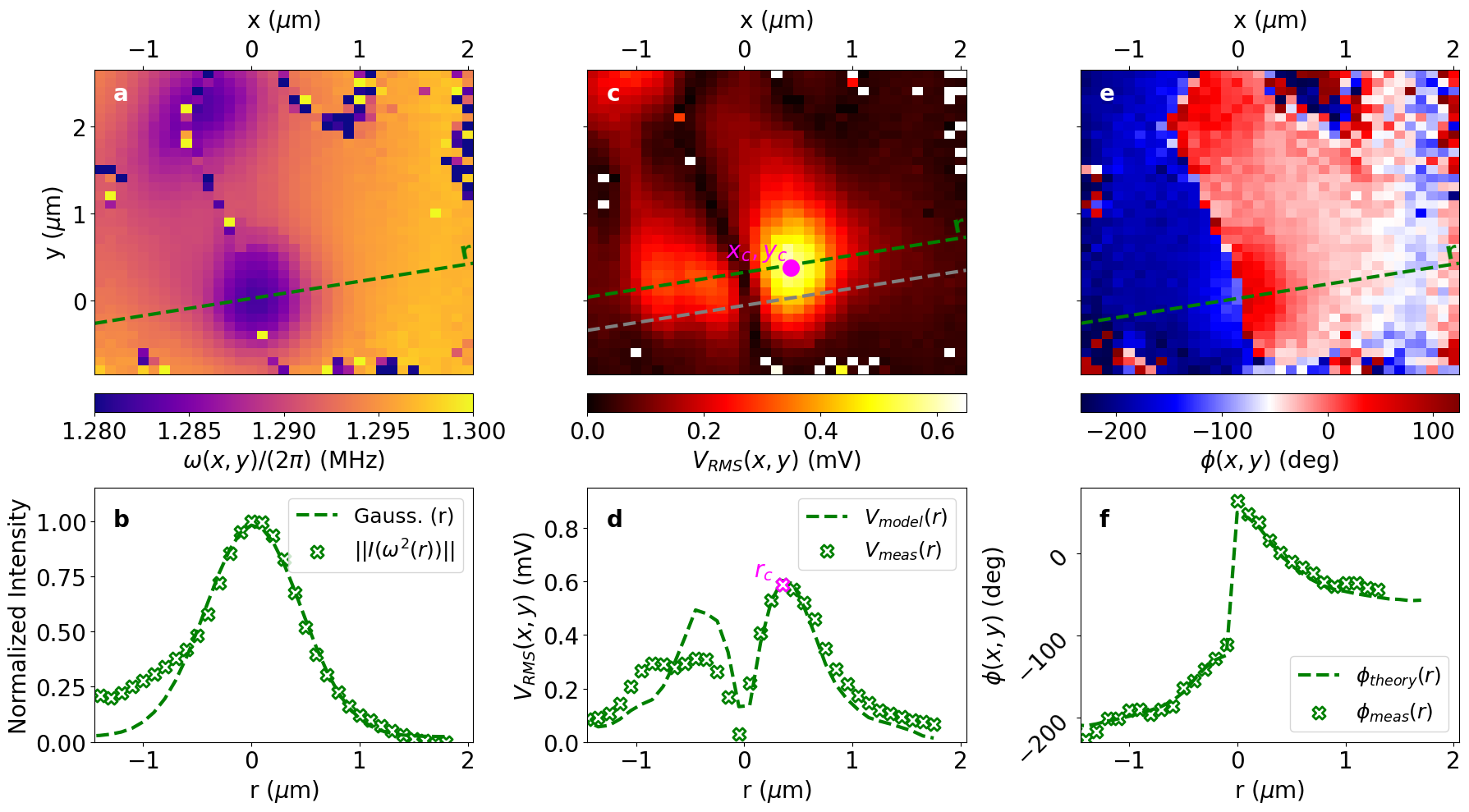}
    \caption{2D maps and line profiles from a 7-hour raster scan at an average beam power of $12.5~\mu$W at 633~nm. Each point corresponds to a parameter extracted from a Lorentzian fit to a frequency sweep at position ($x$,$y$). Calibration to thermomechanical noise was performed at $(x_\mathrm{c}, y_\mathrm{c})$ (or $r_\mathrm{c}$). A green line along the NW motion axis $r$ marks the profile extraction direction. Maps show the spatial dependence of resonance frequency (a), RMS amplitude at resonance (c), and phase response (e), all obtained from driven frequency sweeps. In (c), the dark gray line marks the peak center and the green line traces the signal maximum, offset by a slight azimuthal tilt of the NW. Corresponding line profiles are shown below: (b) normalized beam intensity inferred from the frequency response with a Gaussian fit (dashed), (d) measured versus modeled voltage amplitude with $R(r)=const.$ (see supplementary material) and calibration point $r_\mathrm{c}$, and (f) measured phase fit with a model based on bulk wave dispersion in the substrate.
    }
    \label{fig:MapsNLines}
\end{figure*}


Fig.~\ref{fig:MapsNLines}a shows a map of the resonance frequency when scanned over the NW, resulting in a downward frequency shift due to photothermal heating, along with a nearby nanowire of similar frequency. The shift follows the laser intensity rather than its gradient, indicating static photothermal softening rather than dynamic optical spring effects.\cite{bellon2024temperature}
We extracted the beam profile from the frequency map (Fig.~\ref{fig:MapsNLines}b) and fitted it with a Gaussian of beam waist $\mathrm{w}_0\approx884$~nm, somewhat smaller than the knife-edge value of 992~nm (see supplementary material).

Fig.~\ref{fig:MapsNLines}c shows the measured signal amplitude $V_{\mathrm{RMS}}$, which follows the laser intensity gradient (Fig.~\ref{fig:MapsNLines}d), with an offset of approximately 250~nm from the heating center shown in Fig.~\ref{fig:MapsNLines}c. This offset suggests that the center of scattering differs from the center of absorption, likely due to a slight tilt relative to the optical axis: the Au tip primarily scatters, while the NW body primarily absorbs. Asymmetry between the lobes may arise from spatial variation in the substrate reflectance $R(r)$. The maximal signal was obtained near $r_c\approx\mathsf{w_0}/2$, where the gradient of the Gaussian beam is maximal. This point was used to calibrate the measured signal, according to (\ref{eq:VAC}), with the calibration factor $C(r_c)$ (\ref{eq:Cc}) obtained from the thermomechanical noise measurements.

At other positions $r\neq r_\mathrm{c}$, however, the signal may fall below the transduction noise floor, inhibiting resolution of $S_{V_\mathrm{thm}}$. In any case, $C(r)$ can be related to the calibrated value at $r_\mathrm{c}$ via (see supplementary material)
\begin{equation}
C(r) = C(r_\mathrm{c})\frac{R(r_\mathrm{c})}{R(r)}\frac{\frac{d}{dr}I(r)|_{r_\mathrm{c}}}{\frac{d}{dr}I(r)\phantom{|_{r_\mathrm{c}}}}\,.
\end{equation}
The normalized reflectance $R(r)/R(r_\mathrm{c})$ is then inferred from the ratio of the measured to the modeled signal amplitude in Fig.~\ref{fig:MapsNLines}d, where the model assumes a constant reflectance (dashed line).

In Fig.~\ref{fig:MapsNLines}e, a steady phase variation with laser position arises from a frequency-dependent delay due to bulk wave dispersion in the piezoactuator--substrate system, which becomes significant when driving near the NW resonance frequency. This is confirmed in Fig.~\ref{fig:MapsNLines}f by fitting the phase values along $r$ with a scaled resonance frequency response model (dashed line), including the expected abrupt 180$^\circ$ phase reversal at $r\approx0$.\cite{doster2022observing} This reversal prevented consistent PLL lock throughout the scan. The variation in $Q$ with position is provided in the supplementary material.

\begin{figure*}[t]
    \centering
    \includegraphics[width=\textwidth]{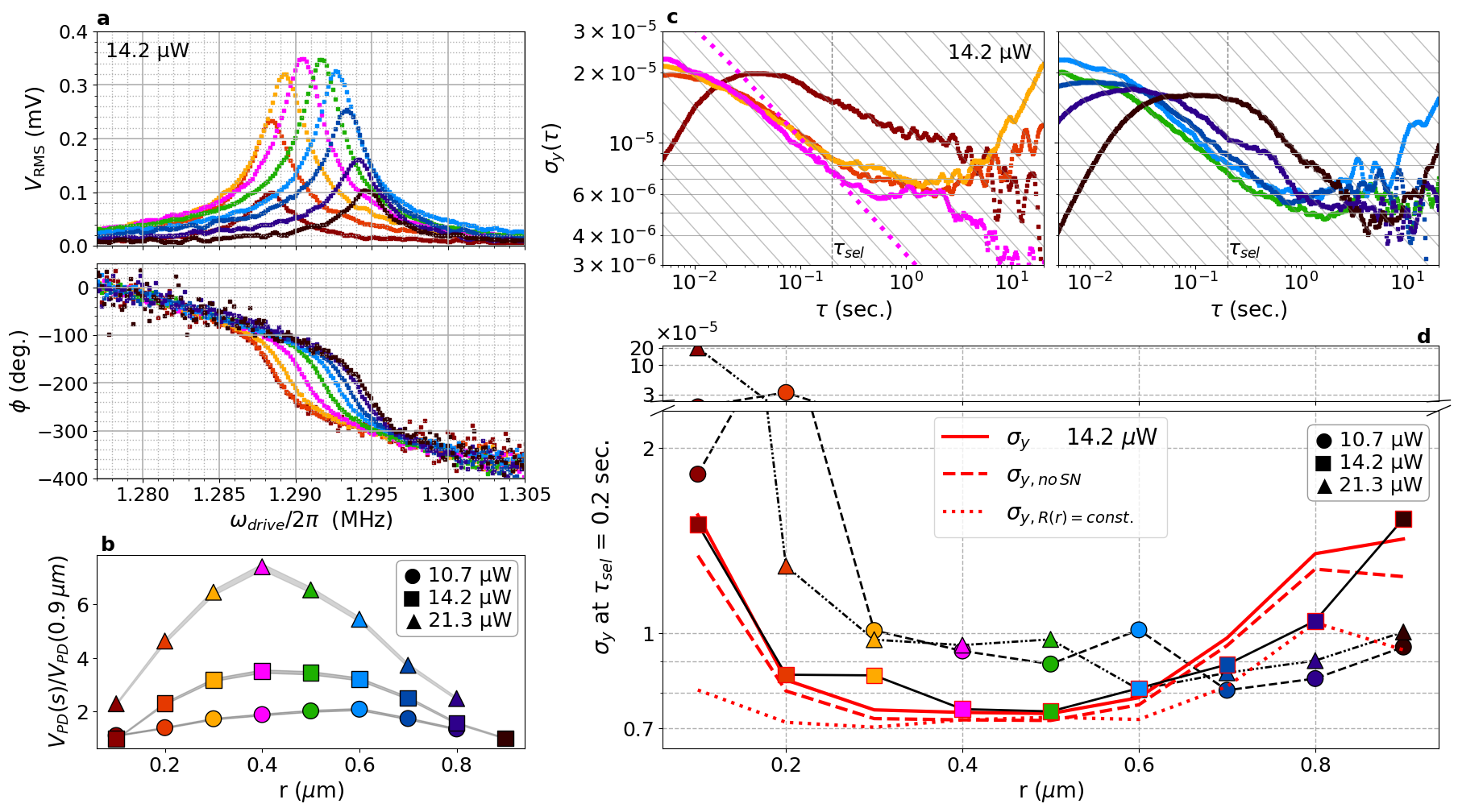}
    \caption{Frequency stability study. Results from line scans along the nanowire axis $r$ from $r = 0.1~\mu$m (dark red) through $r_\mathrm{c}$ and toward $r = 0.9~\mu m$ (black). (a) Mixed-signal amplitude and phase from frequency sweeps at $14.2~\mu$W. (b) Photo diode amplitude, normalized to $r = 0.9~\mu$m of $10.7\mu W$. (c) Allan deviation $\sigma_y(\tau)$ at $14.2\,\mu W$; magenta, dashed line is the expected white noise floor at $r_\mathrm{c}$. Gray lines with $\sqrt{\tau}$ dependence highlight white noise trends; vertical dashed line is chosen integration time $\tau_{\mathrm{sel}} = 0.15$ sec. for (d), the Allan deviation for all line scans. Red lines show the expected theoretical thermomechanical limit (solid), that with shot noise removed (dashed), and that with shot noise but where the reflectance of the substrate is constant (dotted) at $14.2\,\mu W$.
    }
    \label{fig:AdevwithfandAfig}
\end{figure*}

Finally, the frequency stability of the InP NW is studied. 
Figure~\ref{fig:AdevwithfandAfig}a shows the PD voltage amplitude response at each position $r$ (top) and the corresponding phase response (bottom) obtained from drive frequency sweeps at $P_\mathrm{in}=14.2~\mu W$. The frequency shifts down as the laser beam becomes more centered on the NW due to photothermal heating and the mixed signal's amplitude ($V_{\mathrm{RMS}}$) is the highest at the calibration point, $r_\mathrm{c}$, where the NW intersects the steepest gradient of the Gaussian intensity profile. Fig.~\ref{fig:AdevwithfandAfig}b shows the signal strength for three incident powers as a function of position. According to (\ref{eq:VAC}), the signal increases with power and has a maximum typically near the gradient maximum of the Gaussian beam. 

Fig.~\ref{fig:AdevwithfandAfig}c presents the measured Allan deviations at a fixed power of 14.2~µW for varying laser positions. At short integration times, the PLL bandwidth limits transduction noise; at longer times, $1/f$ noise seems to dominate.\cite{bevsic2023schemes} In the intermediate regime, the Allan deviation curves follow an approximate $\tau^{-1/2}$ trend (gray slanted lines), where the sensitivity is limited by white noise: either thermomechanical fluctuations or photon shot noise. The NW's extremely low effective mass ensures that thermomechanical noise dominates over thermal frequency fluctuations by several orders of magnitude.\cite{cleland2002noise} The Allan deviation at 14.2~$\mu$W is minimized near the signal maximum ($r_\mathrm{c}$, magenta). Centering the beam on the NW increases the temperature and dramatically raises thermomechanical noise, which then dominate (dark red line). Allan deviations at other powers are provided in the supplementary material, along with an analysis of the position- and power-dependent optical force, derived from the Lorentzian fit parameters at each power and position.

We use the 14.2~$\mu$W data in Fig.~\ref{fig:AdevwithfandAfig}c to compare with our theoretical prediction. 
In a closed-loop oscillator scheme, the fractional frequency noise PSD due to the additive phase noise is\cite{bevsic2023schemes,schmid2023fundamentals} 
\begin{equation}\label{eq:Sy}
\begin{split}
    S_y(\omega, r) = \frac{1}{2Q^2}\frac{1}{u_\mathrm{dr}^2} \bigg[ 
    & S_{u_\mathrm{thm}}(\omega_0)\, |H_{\varphi_\mathrm{thm}}(\omega)|^2 \\
    +\; & C^2(r)\, S_{V_\mathrm{SN}}\, |H_{\varphi_\mathrm{tr}}(\omega)|^2 
    \bigg]\,,
\end{split}
\end{equation}
where $H_{\varphi_\mathrm{thm}}$ and $H_{\varphi_\mathrm{tr}}$ are the thermomechanical and transduction transfer functions, defined by the resonator time constant and PLL bandwidth.\cite{bevsic2023schemes} 
$S_{V_\mathrm{SN}}$ is the shot-noise PSD at the photodetector, in units of [$\mathrm{V}^2/\mathrm{Hz}$],
\begin{equation}\label{eq:SVsn}
    S_{V_\mathrm{SN}} = 2 \frac{h c}{\lambda} P\, g_{\mathrm{PD}}^2,
\end{equation}
where $h$, $c$, and $\lambda$ are Planck's constant, the speed of light, and the laser wavelength, respectively. The Allan deviation, as measured in Fig.~\ref{fig:AdevwithfandAfig}c, can be calculated from the one-sided fractional frequency noise PSD (\ref{eq:Sy}) by $\sigma_y^2(\tau) = (4/(\pi\tau^2)) \int_{0}^{\infty} S_y(\omega)\sin^4(\omega\tau/2)/\omega^2\, d\omega$, where $\tau$ is the integration time.\cite{rubiola2008phase} The detailed calibration procedure and calculation of $\sigma_y^2(\tau)$, including the use of Fig.~\ref{fig:MapsNLines}d for determining $R(r)$, can be found in the supplementary material. The resulting theoretical Allan deviation, computed over the full range of integration times at $r_\mathrm{c}$, is shown as the magenta dotted line in Fig.~\ref{fig:AdevwithfandAfig}c.

Fig.~\ref{fig:AdevwithfandAfig}d shows a comparison of the Allan deviations at the intermediate time $\tau_\mathrm{sel}=0.15$~s as a function of increasing optical power. The comparison reveals no significant improvement with laser power. This can be explained by excessive NW heating, which increases thermomechanical noise (Eq.~\ref{eq:Suth}) and counteracts the signal-strength enhancement (Eq.~\ref{eq:VAC}) at higher power. The experimental data for one power value is compared with the theoretical model, depicted as the solid red line in Fig.~\ref{fig:AdevwithfandAfig}d. The dashed red line shows the prediction with shot noise removed, demonstrating that thermomechanical noise dominates at these power levels. The dotted red line represents the model predictions neglecting the variations in substrate reflectance.

This work provides a framework for interpreting the photodetector signal in on-axis optical detection of vertical nanowires and for diagnosing the noise contributions that limit frequency stability. We establish a spatially resolved calibration model derived from the probing laser intensity profile, inferred from the nanowire's photothermal frequency response at different incident powers. This calibration relates the nanowire's flexural resonance amplitude and intrinsic thermomechanical noise to the observed signal. Shot noise plays a nontrivial role in the noise budget; however, unique to this transduction scheme, the spatial variation in substrate reflectance contributes even more significantly. Increasing the laser power does not significantly improve frequency stability, because the associated temperature rise enhances thermomechanical noise and counteracts the gain in signal strength. The optimal laser position lies near the maximum of the intensity profile gradient, although the measured Allan deviations at different powers do not uniformly follow this prediction, suggesting that additional position-dependent effects, such as optical back-action or spatially varying dissipation, may contribute at higher powers. Avoiding this trade-off between signal gain and thermal noise will require nanowire materials with lower optical absorption. The model presented here provides a quantitative basis for such optimization and for predicting the detection limits of nanowire-based sensors for mass, force, or photothermal sensing.

\section*{Supplementary Material}
The supplementary material contains derivations and supporting data for the analysis presented here, including the effective mass calculation, the photothermal beam profile reconstruction, the position-dependent calibration procedure, the full fractional frequency noise PSD derivation, and evidence for position- and power-dependent optical forces. Supporting figures show the knife-edge beam waist measurement, raster scan data on a second nanowire, quality factor maps, and Allan deviations at all laser powers.

\section*{Acknowledgements}
This work was supported by the European Research Council under the European Union's Horizon 2020 research and innovation programme (Grant Agreement No.~716087, PLASMECS), the Swedish Research Council, NanoLund through Myfab, and the Wallenberg Initiative Materials Science for Sustainability (WISE) funded by the Knut and Alice Wallenberg Foundation.

\section*{Author Declaration}
\subsection*{Conflict of Interest}
The authors have no conflicts to disclose.

\subsection*{Author Contributions}
R.G.W.: Conceptualization, Methodology, Formal analysis, Investigation, 
Writing -- original draft, Writing -- review \& editing. K.K.: Formal 
analysis, Methodology, Writing -- review \& editing. L.H.: Resources. 
M.B.: Resources, Funding acquisition, Writing -- review \& editing. 
S.S.: Conceptualization, Supervision, Funding acquisition, Writing -- 
review \& editing.

\section*{Data Availability}
The data that support the findings of this study are available from the corresponding author upon reasonable request.

\section*{References}
\nocite{*}
\bibliography{WESTLib}

\end{document}